\documentclass[a4paper,
               biblatex,     
               keeplastbox,   
               ]{jacow}
\usepackage{pdfpages,multirow,ragged2e} %
\makeatletter%
	\ifboolexpr{bool{xetex}}
	 {\renewcommand{\Gin@extensions}{.pdf,%
	                    .png,.jpg,.bmp,.pict,.tif,.psd,.mac,.sga,.tga,.gif,%
	                    .eps,.ps,%
	                    }}{}
\makeatother

\ifboolexpr{bool{xetex} or bool{luatex}} 
 {}                                      
 {\usepackage[utf8]{inputenc}}           

\usepackage[USenglish]{babel}

\ifboolexpr{bool{jacowbiblatex}}%
 {%
  \addbibresource{MOPC71.bib}
 }{}
\begin{document}

\title{Assessing global crabbing scheme feasibility for Electron-Ion Collider\thanks{Work supported by 
Brookhaven Science Associates, LLC under Contract No. DE-SC0012704 with the U.S. Department of Energy}}

\author{Derong Xu\thanks{dxu@bnl.gov}, Yun Luo, Daniel Marx, Christoph Montag, Brookhaven National Laboratory, Upton, NY, USA \\
		Yue Hao, Michigan State University, East Lansing, MI, USA}
	
\maketitle

\begin{abstract}
The Electron-Ion Collider (EIC) plans to utilize the local crabbing crossing scheme. 
This paper explores the feasibility of adopting a single crab cavity with adjusted voltage, 
inspired by the successful global crabbing scheme in KEKB, to restore effective head-on collisions. 
Using weak-strong simulations, the study assesses the potential of this global crabbing scheme for the 
EIC while emphasizing the need for adiabatic cavity ramping to prevent luminosity loss. 
Additionally, the research outlines potential risks associated with beam dynamics in implementing this scheme.
\end{abstract}

\section{Introduction}
A crab cavity is a specialized type of radio-frequency (RF) cavity designed to deliver a time-dependent 
electromagnetic transverse kick to particle bunches. In colliders, this transverse kick tilts the bunches 
at the interaction point (IP), ensuring that beams from two storage rings collide in a head-on manner. 
This method, known as the crab crossing scheme \cite{wei1996crab}, 
compensates for the geometric luminosity loss resulting from a large 
crossing angle, thereby significantly enhancing the collider's luminosity and overall performance.

The crab crossing scheme, along with the concept of crab cavities, was initially introduced for linear 
colliders \cite{palmer1988energy} and subsequently adapted for use in circular colliders \cite{oide1989beam}.
This scheme was first successfully implemented at the KEKB-factory \cite{abe2007compensationof} (High Energy Accelerator 
Research Organization), where a world record luminosity of $2.1\times 10^{34}~\mathrm{cm^{-2}s^{-1}}$ was 
obtained.  Additionally, the experiment demonstrating crabbing with high energy protons was successfully conducted 
at CERN’s Super Proton Synchrotron (SPS) \cite{calaga2021first}. 
Many future particle colliders, including the High Luminosity Large Hadron Collider (HL-LHC) \cite{arduini2016high} and 
the Electron-Ion Collider (EIC) \cite{willeke2021electron}, have incorporated the crab crossing scheme to meet 
their luminosity objectives.

The EIC, to be constructed at Brookhaven National Laboratory (BNL), is designed to collide polarized 
high-energy electron beams and hadron beams. It aims to achieve luminosities up to $10^{34}~\mathrm{cm}^{-2}\mathrm{s}^{-1}$ 
within a center-of-mass energy range of $29-140~\mathrm{GeV}$. 
The EIC will feature two storage rings: the Hadron Storage Ring (HSR) and the Electron Storage Ring (ESR), 
both of which will be accommodated within the existing tunnel of the Relativistic Heavy Ion Collider (RHIC).
A distinctive feature of the EIC is the implementation of crabbing for both electron and hadron beams.

\section{Global vs local crabbing}
\begin{figure}
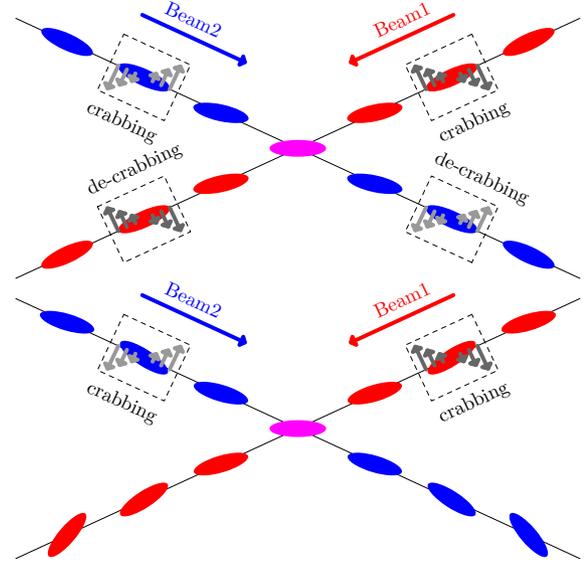

    \centering
    \includegraphics[width=0.9\columnwidth]{MOPC71_f1.1}
    \includegraphics[width=0.9\columnwidth]{MOPC71_f1.2}
    \caption{Schematic of the crab crossing scheme: top (local), bottom (global).}
    \label{fig:crabbing}
\end{figure}

There are two distinct crab crossing schemes: a local scheme and a global scheme.
Figure \ref{fig:crabbing} illustrates the distinctions between these crab crossing schemes.

In the local scheme, a pair of crab cavity sections are strategically positioned: one upstream and one 
downstream of the IP. The upstream crab cavity induces a tilt in the bunch, which is subsequently cancelled by 
the downstream crab cavity. Consequently, the bunch is only tilted within these crab cavities, 
resulting in no tilt beyond this specific area.

Conversely, the global scheme employs crab cavities located at one specific point in the accelerator. 
The bunch tilt is never cancelled, but evolves throughout the entire ring, intertwining the horizontal 
and longitudinal dynamics.

The concept of crab dispersion is introduced to quantify the bunch tilt effect induced by the 
crab cavities \cite{PhysRevAccelBeams.25.071002}. 
In both the local and global schemes, achieving an effective head-on collision necessitates specific crab dispersion at the
IP. This required crab dispersion can be mathematically represented as follows:
\begin{equation}
    \boldsymbol\zeta^*=\left(\theta_c,0\right)
    \label{eq:zetaStar}
\end{equation}
where $\theta_c$ is the half crossing angle. 
To meet the condition specified in Eq. (\ref{eq:zetaStar}), two adjustment mechanisms, or ``knobs'', are essential.

In the local scheme, the available knobs are the upstream and downstream crab cavity voltages.
Two more knobs are needed to cancel out the crab dispersion.
\begin{equation}
\begin{gathered}
    |V_{b,a}|=\frac{cE}{2\pi f_\mathrm{b,a}\sqrt{\beta^*\beta_{b,a}}}\theta_c\\
    \Psi_b=\frac{\pi}{2}\ \mathrm{mod}\ \pi,\quad
    \Psi_a=\frac{\pi}{2}\ \mathrm{mod}\ \pi
    \label{eq:local}
\end{gathered}
\end{equation}
where the subscript $b$ refers to parameters associated with the crab cavity positioned before the IP, 
while $a$ denotes parameters for the cavity located after the IP. 
The variable $V$ is the required crab cavity voltage, $c$ speed of the light, 
$E$ the particle energy, $f$ the crab cavity frequency, 
while $\beta^*$ and $\beta_{b,a}$ refer to the beta functions at the IP 
and at the respective crab cavity locations.
$\Psi_{b}$ ($\Psi_a$) indicates the phase advance needed from the crab cavity (IP) and the IP (crab cavity).

In the global scheme, control over the beam dynamics is achieved using two primary variables: the crab cavity voltage 
and the phase advance between the crab cavity and the IP. 
The equations defining these control parameters are:
\begin{equation}
\begin{gathered}
    |V_c|=\frac{cE}{2\pi f_c\sqrt{\beta^*\beta_c}}\cdot 2\theta_c\sin\left(\pi\nu\right)\\
    \Psi_c=\pi\nu\ \mathrm{mod}\ \pi
    \label{eq:global}
\end{gathered}
\end{equation}
where $\nu$ is the tune in the crabbing plane when the crab cavities are turned off and 
in the absence of beam-beam interactions.

KEKB was the world's first collider to use a crab crossing scheme, choosing the global method \cite{funakoshi2014operational}. 
For the HL-LHC, both local and global schemes were considered \cite{PhysRevSTAB.12.101002}, 
but the global scheme did not fit LHC's varied crossing angles at IPs. 
The HL-LHC upgrade thus adopted the local scheme for better performance and machine protection \cite{calaga2018crab}. 
The EIC will also use the local scheme from the start, but revisiting the global scheme presents an opportunity to 
explore alternative strategies.

\section{Global crabbing in HSR}
In the case of the ESR, the global crabbing scheme has been ruled out. 
The latest working point for the ESR has tunes set to $(0.08,0.14,-0.069)$ based on the beam-beam study 
and polarization requirement \cite{luo:ipac2023-mopa047,osti_1837215,ESR2024}. 
The negative sign signifies that, in the longitudinal 
phase space, particles rotate in a direction opposite to that in the horizontal or vertical phase space.
Because $\nu_x$ is very close to $\nu_z$, even a minor crabbing leakage can cause instability in the horizontal 
plane \cite{PhysRevAccelBeams.25.071002}. To completely eliminate crabbing leakage, the downstream crab cavity is 
positioned further away from the IP, ensuring that the phase advance between the upstream and downstream crab cavities 
is precisely $2\pi$. This approach has been implemented in the design of the ESR lattice, 
as documented in \cite{xu:ipac2022-wepopt050}.

For the HSR, the working point is set to $(0.228,0.210,-0.010)$ with the transverse tunes significantly distanced from the 
longitudinal tune. This separation enables the implementation of the global crabbing scheme within the HSR.

We use a custom weak-strong code to check the feasibility of a global crabbing scheme in the HSR.
In the simulation, the strong electron beam remains a rigid Gaussian distribution, and is cut into multiple slices.
The weak proton bunches are simulated by a number of macro particles.
A second-order harmonic crab cavity is used \cite{xu:ipac2021-wepab009}.
The one-turn map at IP is represented by the linear betatron map.
The beam-beam kick is calculated by the Bassetti and Erskine formula \cite{bassetti1980closed}.

\begin{figure}[!htbp]
    \centering
    \includegraphics[width=0.9\columnwidth]{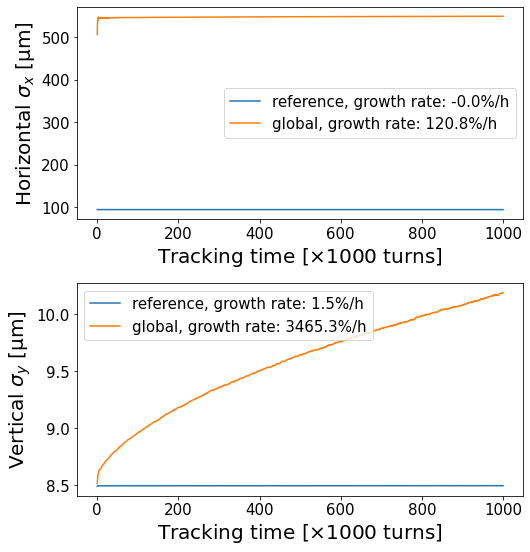}
    \caption{Comparison of weak-strong simulations for local and global crabbing schemes: horizontal beam
    size evolution (top), vertical beam size evolution (bottom).}
    \label{fig:globalHSR}
\end{figure}

Figure \ref{fig:globalHSR} presents a comparison of the results from weak-strong simulations. 
The blue and orange curves illustrate the outcomes for the local and global crabbing schemes, respectively, 
with crab cavity voltages adjusted in accordance with Equations (\ref{eq:local}-\ref{eq:global}). 
The blue curve serves as a benchmark for comparison. In the case of the orange curve, 
the horizontal dimension significantly increases upon reaching its ``equilibrium'', and the growth rate of the 
vertical dimension is notably quicker. This phenomenon is attributed to the beam's initial Gaussian distribution 
and the absence of horizontal-longitudinal coupling, leading to a beam distribution that does not align with 
the lattice configuration.

\begin{figure}[!htbp]
    \centering
    \includegraphics[width=0.9\columnwidth]{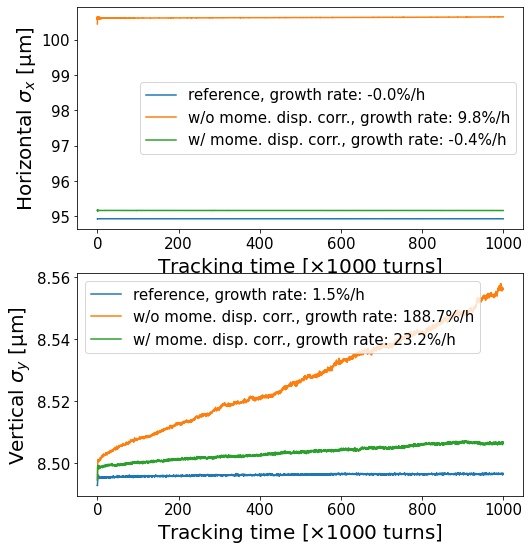}
    \caption{Tracking with initial matched beam distribution: horizontal beam
    size evolution (top), vertical beam size evolution (bottom). 
    For the orange curve, momentum dispersion has been optimized to ensure complete decoupling of the 
    linear one-turn map in the head-on frame, separating transverse and longitudinal planes.}
    \label{fig:matched}
\end{figure}

To validate the effects of the initial mismatched beam distribution, 
Figure~\ref{fig:matched} presents the tracking outcomes when the initial beam is 
aligned with the lattice. This adjustment leads to a diminished horizontal beam size 
and a reduction in vertical expansion, as opposed to Fig.~\ref{fig:globalHSR}. 
Besides the initial beam distribution, the nonzero longitudinal tune also lead to linear synchro-betatron 
coupling when the crab cavity is turned on, as detailed in the study \cite{PhysRevAccelBeams.25.071002}. 
A potential strategy involves altering the momentum dispersion at the IP to completely decouple the 
linear one-turn map in the head-on frame. As shown in Fig.~\ref{fig:matched}, the green curve shows
much slower growth in vertical plane compared with the orange curve. 

\begin{figure}[!htbp]
    \centering
    \includegraphics[width=0.9\columnwidth]{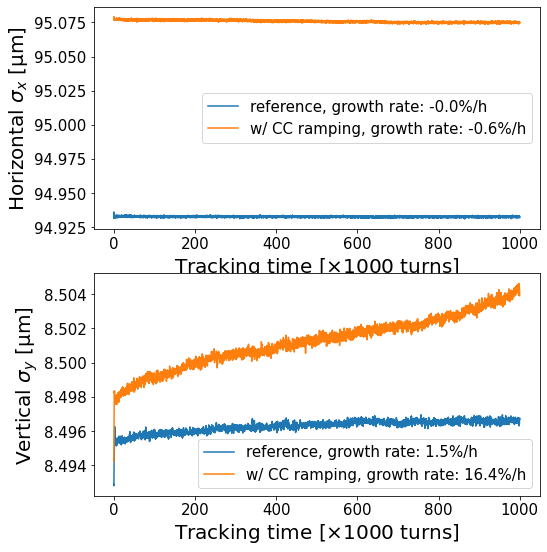}
    \caption{Tracking with the crab cavity voltage ramped over $1000$ turns: horizontal beam
    size evolution (top), vertical beam size evolution (bottom).}
    \label{fig:ramped}
\end{figure}

To create an initial matched beam distribution, one effective method is to gradually increase the 
crab cavity's voltage. Initially, the injected beam is perfectly matched to the lattice 
configuration when the crab cavities are deactivated. Subsequently, the voltage of the crab 
cavity is incrementally raised to its intended design values. It's important to note that 
during this gradual ramping process of the crab cavity, no beam-beam interactions occur, 
allowing the beam distribution to adapt smoothly. Once the voltage ramping is complete, 
collisions are initiated. 
The effectiveness of this method is demonstrated in Figure~\ref{fig:ramped}, which displays 
simulation outcomes when the crab cavity's voltage is methodically ramped up across $1000$ turns. 
Notably, the orange curve in this figure mirrors the vertical growth rate seen in the green 
curve of Fig.~\ref{fig:matched}, suggesting that adiabatic ramping effectively achieves a 
matched beam distribution.

\section{Discussion}

It should be emphasized that, although the initial beam aligns with the lattice and the momentum 
dispersion is finely tuned at the IP, vertical growth remains an issue when compared to the 
reference curve, as illustrated in both Fig.\ref{fig:matched} and Fig.\ref{fig:ramped}. In 
the design of the HSR, the absence of efficient cooling techniques necessitates minimizing vertical 
growth to ensure an adequate luminosity lifetime. From this perspective, 
the global crabbing scheme does not reach the level of performance attainable 
with the local crabbing scheme.

The primary incentive for adopting the global scheme is the potential for cost reduction associated 
with crab cavities. In contrast to electron storage rings, hadron beams exhibit significantly 
higher rigidity, necessitating the use of multiple cavities to achieve an effective crab kick 
within the hadron storage ring. The quantity of crab cavities required is directly proportional 
to the voltage needed. An examination of the formula presented in Eq.~(\ref{eq:global}) reveals 
that the requisite voltage is proportional to $2\sin(\pi\nu)$. Applying this to the horizontal 
tune $\nu_x=0.228$ yields a value of $1.3$. Consequently, the reduction in voltage -- 
or equivalently, the number of crab cavities -- is approximately $35\%$. 
However, when considering the second-order harmonic crab cavities, 
the cost savings are not as substantial as those observed in the KEKB project.

The marginal cost savings offered by the global crabbing scheme are outweighed by the 
introduction of synchro-betatron coupling throughout the ring, which in turn presents 
numerous dynamic challenges, including issues related to dynamic aperture, intra-beam scattering, 
and collimation, among others. Given these associated risks, the implementation of the 
global scheme is not recommended for the HSR design.


\section{ACKNOWLEDGEMENTS}
The authors wish to extend their sincere gratitude to Dr. Katsunobu Oide for his suggestion to 
explore a non-closed crabbing bump scheme for the EIC. 

%
%
\ifboolexpr{bool{jacowbiblatex}}%
	{\printbibliography}%
	{%
	
	
} 
%
%


\end{document}